\documentclass[11pt]{article}

\usepackage{amsmath,amssymb}

\makeatletter

 \@addtoreset{equation}{section}
\makeatother

\newcommand {\beq}{\begin{eqnarray}}
\newcommand {\eeq}{\end{eqnarray}}
\newcommand {\non}{\nonumber\\}
\newcommand {\1}[1]{\frac{1}{#1}}

\newcommand {\ph}{\varphi}

\newcommand {\sig}{\sigma}

\newcommand {\dagg}{^{\dagger}}
\newcommand {\lam}{\lambda}
\newcommand {\tr}{{\rm tr}\,}
\newcommand {\GC}{G^{\bf C}}
\newcommand{\vs}[1]{\vspace{#1 mm}}

\newcommand{\om}{\omega}
\newcommand{\Om}{\Omega}
\newcommand{\R}{{\cal R}}
\newcommand{\scale}[3]{\genfrac{(}{)}{0pt}{}{\,#1\;#2\,}{#3}}
\newcommand{\half}{\frac{1}{2}}
\newcommand{\mbf}{\boldsymbol}

\newcommand{\wt}{\widetilde}
\newcommand{\wh}{\widehat}
\newcommand{\ol}{\overline}

\newcommand{\kahler}{K\"{a}hler }

\newcommand{\bsubeq}{\begin{subequations}}
\newcommand{\esubeq}{\end{subequations}}

\newcommand{\w}{\wedge}

\newcommand{\Pd}[2]{\frac{\partial{#1}}{\partial{#2}}}

\newcommand{\ltd}[1]{\frac{d}{d{#1}}}

\newcommand{\del}{\partial}
\newcommand{\il}{e}
\newcommand{\jl}{f}
\newcommand{\kl}{g}

\begin{document}

\allowdisplaybreaks{

\setcounter{page}{0}

\begin{titlepage}

{\normalsize
\begin{flushright}
OU-HET 405 \\
PURD-TH-02-01 \\
{\tt hep-th/0202064} \\
February 2002
\end{flushright}
}

\vs{2}

\begin{center}
{\Huge Calabi-Yau Manifolds of Cohomogeneity One

as Complex Line Bundles}

\vs{7}

\bigskip
{\renewcommand{\thefootnote}{\fnsymbol{footnote}}
{\Large\bf Kiyoshi Higashijima$^1$\footnote{
     E-mail: {\tt higashij@phys.sci.osaka-u.ac.jp}},
 Tetsuji Kimura$^1$\footnote{
     E-mail: {\tt t-kimura@het.phys.sci.osaka-u.ac.jp}}
 and Muneto Nitta$^2$\footnote{
     E-mail: {\tt nitta@physics.purdue.edu}}
}}

\vs{2}

{\large\sl
$^1$ Department of Physics,
Graduate School of Science, Osaka University, \\
Toyonaka, Osaka 560-0043, Japan \\

$^2$ Department of Physics, Purdue University, West Lafayette, IN
47907-1396, USA
}

\end{center}

\setcounter{footnote}{0}


\vs{7}

\begin{abstract}

We present a simple derivation of
the Ricci-flat \kahler metric and
its \kahler potential on the canonical line bundle
over arbitrary \kahler coset space
equipped with the K\"{a}hler-Einstein metric.

\end{abstract}

\end{titlepage}

\section{Introduction}

A lot of attention has been paid to compact Calabi-Yau manifolds,
since they give candidates of
compactification of superstring theory~\cite{CHSW}.
Calabi-Yau manifolds are defined as
K\"ahler manifolds satisfying
the Einstein equation with zero cosmological constant,
namely the Ricci-flat condition.
It is, however, quite difficult to obtain explicit metrics
of Calabi-Yau manifolds in general, since the Ricci-flat condition
is a highly non-trivial partial differential equation.
No explicit metric is known for compact Calabi-Yau manifolds.
We would like to discuss {\it non-compact} Calabi-Yau manifolds.

Isometry of the manifold plays a crucial role
to solve the Einstein equation,
since it often reduces the equation
to an ordinary differential equation
(or a set of algebraic equations).
When the isometry group $G$ is transitive,
the manifold is called homogeneous and can be
written as a coset space $G/H$,
where $H$ is the isotropy group.
However coset spaces cannot have Ricci-flat metric,
since their scalar curvature is positive.
When the isometry is not transitive
but generic orbits have
dimensions less than the whole space by one,
the manifold is called cohomogeneity one
and can be written as ${\bf R} \times G/H$ at least locally.
Hyper-K\"ahler manifolds, which belong to a class of
Calabi-Yau manifolds, of cohomogeneity one were
completely classified by Dancer and Swann~\cite{DS}:
the only one manifold is the Calabi metric~\cite{Ca}
on the cotangent bundle over
the projective space ${\bf C}P^{N-1} =
SU(N)/[SU(N-1) \times U(1)]$,
$T^* {\bf C}P^{N-1}$.
Cohomogeneity one manifolds with holonomy of
$Spin(7)$ or $G_2$ have been recently studied extensively
by many authors
in connection with the compactification
of string theory or M-theory (see, for example, \cite{Spin7,M}).

General discussion of
Calabi-Yau manifolds of cohomogeneity one
has been recently given by Dancer and Wang~\cite{DW}.
A class of Calabi-Yau manifolds of cohomogeneity one
is given by the cotangent bundles over
rank one symmetric spaces, $T^*(G/H)$,
obtained by Stenzel~\cite{St}.
In particular, the cotangent bundle over
$S^{N-1} = SO(N)/SO(N-1)$,
gives the higher dimensional generalization~\cite{CGLP,HKN1}
of the Eguchi-Hanson space~\cite{EH}
and the deformed conifold~\cite{conifold}.
An another important class of the Calabi-Yau manifolds
of cohomogeneity one is given by
the complex line bundles over coset spaces with
K\"{a}hler-Einstein metrics.
They were firstly constructed by Page and Pope~\cite{PP}
in real coordinates.
In connection with supersymmetric nonlinear
sigma models~\cite{Zu} on Calabi-Yau manifolds,
it is important to give such manifolds in complex coordinates
for the manifest supersymmetry on the string world sheet.
This has been done in a series of the papers
given by the present authors~\cite{HKN2,HKN3,HKN4}
for the canonical line bundles over
the Hermitian symmetric spaces,
using the gauge theoretical (K\"ahler quotient) construction
of the base spaces~\cite{HN}.
(See also \cite{PV}.)
A new way of a resolution of
the conical singularity of the conifold is given in~\cite{HKN2},
which is achieved by the line bundle over
the quadric surface $SO(N)/[SO(N-2)\times U(1)]$.
The generalization of the conifold to the one
with the isometry of $E_6$ or $E_7$ is given in~\cite{HKN4},
which is identified with the line bundle over
$E_6/[SO(10) \times U(1)]$ or $E_7/[E_6 \times U(1)]$.

In this paper, we give the Calabi-Yau metric
and its \kahler potential on the canonical line bundle over
an arbitary K\"{a}hler-Einstein manifold of a coset space $G/H$,
using the method of the supersymmetric nonlinear
realization developed in~\cite{BKMU}.
We extensively use the K\"{a}hler-Einstein structures
of the base manifolds
to calculate the determinants of the metrics.

This paper is organized as follows.
In section \ref{KE-metric},
we discuss the \kahler coset spaces $G/H$
and their Einstein metrics.
The useful formula for the determinant of
the metric is given.
In section \ref{expression},
we present the explicit expressions of
the Ricci-flat \kahler metric and its \kahler potential
on the canonical line bundle over
an arbitrary \kahler coset space
endowed with the K\"{a}hler-Einstein metric.
Some examples are given in section \ref{examples}.
Section \ref{discussion} is devoted to discussion.
In Appendix \ref{tensors},
we construct the Ricci tensors of the Riemann and \kahler coset spaces
$G/H$ in terms of the structure constants of $G$.
In Appendix B, we demonstrate
that the Einstein condition on
\kahler coset spaces reduces to algebraic equations
using some examples.

\section{K\"{a}hler-Einstein Metrics on \kahler Coset Spaces}
\label{KE-metric}

A coset space $G/H$ has the unique \kahler metric up to
scale constants~\cite{Bo},
when the isotropy $H$ is in the form of
\begin{align}
 H \ &= \ H_{\rm ss.} \times U(1)^k \; , \label{H}
\end{align}
where $H_{\rm ss.}$ is the semi-simple subgroup of $H$,
and $k \equiv {\rm rank}\, G  - {\rm rank}\, H_{\rm ss.}$
is the dimension of the torus in $H$.
All of the \kahler coset spaces and their complex structures
were classified by Bordemann, Forger and R\"{o}mer~\cite{BFR}
in terms of the painted Dynkin diagram.
The \kahler potential of arbitrary \kahler coset space was
given by Itoh, Kugo and Kunitomo~\cite{IKK},
using the supersymmetric nonlinear realization~\cite{BKMU}.
We briefly discuss their construction
of \kahler potentials in this section.

We denote the Lie algebra of $G$ by its Calligraphic font
${\cal G}$.
The generators of ${\cal G}$ can be divided
into those of ${\cal H}$ and their orthonormal complements:
$\{S_{\wh{a}} \} = {\cal H}$ ($\wh{a}=1,\cdots,\dim H$)
and $\{X_{\wh{I}} \} = {\cal G} - {\cal H}$ ($\wh{I} =1,\cdots, \dim
G/H$),
with $\tr (S_{\wh{a}} X_{\wh{I}}) = 0$.
A complex structure on $G/H$ can be defined
by dividing $X_{\wh{I}}$ into two sets of non-Hermitian generators,
$X_I$ and $X_{I^*}$,
and by regarding the complex isotropy algebra and its complement
as ${\cal H}^{\bf C} \oplus \{ X_I \} = \wh{\cal H}$ and
$\{ X_{I^*} \} = {\cal G}^{\bf C} -\wh{\cal H}$, respectively.
Under these definitions, there exists a homeomorphism between
real and complex coset spaces:
$G/H \simeq \GC/\wh{H}$.
The representative of the complex coset space $\GC/\wh{H}$
is given by
\begin{align}
 \xi(\ph) \ &= \ \exp (i \ph^i X_{I^*} \delta^I_i ) \; ,
\end{align}
in which the complex coordinates $\ph^i$ parametrize $\GC/\wh H$
($i = 1,\cdots, \dim G/H$).
Here we have used the matrices of
the fundamental representation of $G$ for the generators,
but we do not write it explicitly.
The transformation law of the coordinates under
$g \in G$ is given by
\begin{align}
 \xi \ \to \ \xi' \ &= \ g \xi \wh{h}'{}^{-1}(g,\xi) \; , \ \ \
 \wh h' \ \in \ \wh{H} \; ,
\end{align}
where $\xi' =  \exp (i \ph'{}^i X_{I^*} \delta^I_{i})$
and $\wh{h}'$ is needed to project $g \xi$ onto the coset representative.
With prepareing $k$ projection matrices
$\eta_{\alpha}$ ($\alpha=1,\cdots,k$),
satisfying the projection condition~\cite{BKMU}
\begin{align}
 \eta \wh{H} \eta \ &= \ \wh{H} \eta \; , \ \ \
 \eta^2 \ = \ \eta \; , \ \ \
 \eta^{\dagger} \ = \ \eta \; ,
\end{align}
the \kahler potential of $G/H$ is given by~\cite{IKK}
\begin{align}
 \Psi (\ph,\ph^*)
 \ &= \ \sum_{\alpha=1}^{k} v_{\alpha}
   \log \det{}_{\eta_{\alpha}} \xi\dagg\xi  \;,
\end{align}
where $\det_{\eta}$ denotes the determinant of
the subspace projected by $\eta$,
and $v_{\alpha}$ are real positive constants.
This transforms under $g \in G$ by
\begin{align}
 \Psi \ \to \ \Psi'
 \ &= \ \Psi + \gamma(\ph) + \gamma^*(\ph^*) \; , \ \ \
 \gamma(\ph) \equiv \sum_{\alpha=1}^{k} v_{\alpha}
  \log \det{}_{\eta_{\alpha}} \wh{h}'{}^{-1}(g,\xi) \; .
\end{align}

The \kahler metric is given by
$g_{ij^*} = \del_i \del_{j^*} \Psi$,
where $\del_i$ denotes differentiation with respect to $\ph^i$.
With a suitable choice of $v_{\alpha}$,
any \kahler coset space $G/H$ becomes Einstein \cite{On,BFR,Be}
(see Appendices A and B):
\begin{align}
 \R_{ij^*} \ &= \ h g_{ij^*} \; , \label{Einstein}
\end{align}
where $\R_{ij^*} \equiv
- g^{kl^*} R_{kl^*ij^*}$
is the Ricci-tensor with $R_{ij^*kl^*}$
being the curvature tensor,
and $h$ is a real positive constant.
Using an another expression of the Ricci tensor, we obtain
\begin{align}
\R_{ij^*} \ &= \ - \del_i \del_{j^*} \log \det g_{kl^*}
 \ = \ h \del_i \del_{j^*} \Psi \; .
\end{align}
This can be integrated to give the determinant formula~\cite{Ca},
\begin{align}
 \det g_{ij^*}
 \ &= \ e^{ - h \Psi} |{\rm hol.}|^2 \; ,
  \label{det-formula}\end{align}
where ``hol.'' denotes a holomorphic function.
We extensively use this formula in the next section.\footnote{
In our previous papers \cite{HKN2,HKN3,HKN4},
we used complex isotropy transformations,
$g_{ij^*} \to g'_{ij^*} = (\wh{h} g \wh{h}^{\dagger})_{ij^*}$
to evaluate determinants of metrics of hermitian symmetric spaces. In
some cases, the coordinates $\ph^i$ can be brought to a single component
while this is not generally possible. The use the determinant formula
(\ref{det-formula}) improves this flaw and offers
the far simpler method.
}

\section{Ricci-flat Metrics and \kahler Potentials
on Line Bundles} \label{expression}

In this section we construct a Ricci-flat \kahler metric
and its \kahler potential on canonical line bundle
over an arbitrary \kahler coset space as a base manifold.
We consider a direct product of $N$ \kahler
coset spaces $G_a/H_a$ ($a=1,\cdots,N$) as a base manifold:
\begin{align}
 M \ &= \ (G_1/H_1) \times (G_2/H_2) \times \cdots \times (G_N/H_N) \; ,
\end{align}
and we label each quantity in the last section by the index $a$.
The isometry of $M$ is $G \equiv \prod_{a=1}^N G_a$.
The \kahler potential for the invariant metric on
$M$ is given by
\begin{align}
 \Psi \ &= \ \sum_{a=1}^N \Psi_a \; , \ \ \
 \Psi_a (\ph,\ph^*)
 \ = \ \sum_{\alpha_a=1}^{k_a} v_{\alpha_a}
 \log \det{}_{\eta_{\alpha_a}} \xi_a^{\dagger} \xi_a  ,
\end{align}
where $k_a$ is the dimension of the torus in $H_a$.
This transforms under the isometry
$g = (g_1,\cdots,g_N) \in G$ by
\begin{align}
 \Psi \ \to \ \Psi'
\ &= \ \Psi + \sum_{a=1}^N
 \big[ \gamma_a(\ph) + \gamma_a^*(\ph^*) \big] \; , \ \ \
 \gamma_a(\ph) \equiv \sum_{\alpha_a=1}^{k_a} v_{\alpha_a}
  \log \det{}_{\eta_{\alpha_a}} \wh{h}'{}_a^{-1} \; . \label{G-tr-K}
\end{align}
We introduce a fiber $\sig$ of a complex line bundle over $M$
whose transformation law under $g \in G$ is defined by
\begin{align}
 \sig \ \to \ \sig'
 \ &= \ \exp \Big(- \sum_{a=1}^N
       h_a \gamma_a (\ph) \Big) \sig \; ,\label{sigma}
\end{align}
where $\gamma_a(\ph)$ is defined in (\ref{G-tr-K}).
This transformation law is the same with the one of \cite{Nib}
except for a factor of $h_a$.
The reason of the inclusion of $h_a$ in the transformation law
is clarified below.
In \cite{Nib}, the consistency of the global definition of
the manifold is discussed.
The $G$-invariant is found as
\begin{align}
 X \ &\equiv \ \log \big( |\sig|^2 e^{ \sum_{a=1}^N h_a \Psi_a} \big)
   \ = \ \log |\sig|^2 + \sum_{a=1}^N h_a \Psi_a
   \ = \ \log |\sig|^2 + \wh{\Psi} \; , \label{X}
\end{align}
where $\wh{\Psi} \equiv \sum_{a=1}^N h_a \Psi_a$ can be regarded as
a potential for the Ricci tensor.
We assume the \kahler potential of the line bundle over $M$
as a function of $X$:
\begin{align}
 {\cal K} \ &= \ {\cal K}(X) \; .
\end{align}
We write the coordinates for the total spase
as $z^{\mu} = \{\sig,\ph^i_a\}$,
in which the index $i$ runs through the dimension of $M_a$
for each factor of $M_a$,
but we often omit $a$ if there is no confusion.
Components of the \kahler metric
$g_{\mu \nu^*} = \del_{\mu}\del_{\nu^*} {\cal K}$
are given by
\begin{align}
 g_{\mu \nu^*} \ &= \ \left(
 \begin{array}{cc}
 g_{\sigma \sigma^*} & g_{\sigma j^*} \\
 g_{i \sigma^*} & g_{ij^*}
 \end{array} \right) \; ,
\end{align}
in which each block can be written as
\begin{align}
 g_{\sigma \sigma^*}
\ &= \
{\cal K}'' \Pd{X}{\sigma} \Pd{X}{\sigma^*} \; , \ \ \
g_{\sigma j^*}
\ = \
{\cal K}'' \Pd{X}{\sigma} \Pd{X}{\ph^{*j}} \; , \ \ \
g_{ij^*}
\ = \
{\cal K}'' \Pd{X}{\ph^i} \Pd{X}{\ph^{*j}}
+ {\cal K}' \frac{\del^2 X}{\del \ph^i \del \ph^{*j}} \; ,
\end{align}
where the prime denotes the differentiation with respect to
the argument $X$.
Using equations
$\del X / \del \sigma = 1/ \sigma$ ($\sigma \neq 0$) and
\begin{align}
\frac{\del^2 X}{\del \ph^i \del \ph^{*j}}
\ &= \ \left(
\begin{array}{ccc}
   h_1 g^1_{ij^*} &        & {\bf 0} \\
                  & \ddots & \\
          {\bf 0} &        & h_N g^N_{ij^*}
\end{array} \right) \; ,
\end{align}
the determinant of the metric can be calculated, to give
\begin{align}
 \det g_{\mu \nu^*}
 \ &= \ g_{\sig \sig^*} \cdot
   \det (g_{ij^*} - g_{\sig\sig^*}^{-1} g_{i \sig^*}g_{\sig j^*})
 \ = \
 \frac{1}{|\sigma|^2} {\cal K}''
 ({\cal K}')^d \cdot  \prod_{a=1}^N \det (h_a g^a_{ij^*}) \; ,
 \label{metric-determinant}
\end{align}
where $d \equiv \dim_{\bf C} M$.
Using the determinant formula (\ref{det-formula}) for each $M_a$,
we obtain
\begin{align}
 \det g_{\mu \nu^*}
 \ &= \
  e^{- X} {\cal K}'' ({\cal K}')^{d} |{\rm hol.}|^2 \; .
 \label{det}
\end{align}
Since the Ricci tensor is defined
by $\R_{\mu \nu^*} = - \del_{\mu}
\del_{\nu^*} \log \det g_{\kappa \lam^*}$,
the Ricci-flat condition $\R_{\mu \nu^*} = 0$
implies
\begin{align}
 \det g_{\mu \nu^*} \ &= \ \mbox{(constant)}
 \times |{\rm hol.}|^2 \; .
 \label{RF-condition}
\end{align}
We thus obtain the Ricci-flat condition for
the line bundles over $M$ as
an ordinary differential equation:
\begin{align}
e^{-X}  \ltd{X} ({\cal K}')^D \ &= \ a \; , \label{ODE}
\end{align}
where $a$ is a constant, and $D \equiv d+1$
is complex dimension of the total space.
The solution for ${\cal K}'$ can be obtained as
\begin{align}
{\cal K}' \ &= \ \big( \lam e^X + b \big)^{\frac{1}{D}} \; ,
\label{sol-RF-Calabi}
\end{align}
where $\lam$ is a constant related to $a$ and $D$,
and $b$ is an integration constant.
The \kahler potential can be integrated, to yield
\begin{align}
{\cal K} (X)
\ &= \
D \big( \lam e^X + b \big)^{\frac{1}{D}}
+ b^{\frac{1}{D}}
\cdot I \big( b^{- \frac{1}{D}}
\big( \lam e^X + b \big)^{\frac{1}{D}} ; D \big) \; ,
 \label{kahler-sol}
\end{align}
where the function $I(y;n)$ is defined by
\begin{align}
I (y; n)
\ \equiv \ \int^{y} \! \frac{dt}{t^n - 1}
\ &= \
\frac{1}{n} \Big[ \log \big( y - 1 \big)
    - \frac{1 + (-1)^n}{2} \log \big( y + 1 \big) \Big] \nonumber \\
& \ \ \ \
 + \frac{1}{n} \sum_{r=1}^{[\frac{n-1}{2}]} \cos \frac{2 r \pi}{n}
\cdot \log \Big( y{}^2 - 2 y \cos \frac{2 r \pi}{n} + 1 \Big)
\nonumber \\
\ & \ \ \ \
+ \frac{2}{n} \sum_{r=1}^{[\frac{n-1}{2}]}
\sin \frac{2 r \pi}{n}
\cdot \arctan
\Big[ \frac{\cos (2 r \pi / n) - y} {\sin (2 r \pi /n) } \Big]
\; . \label{fn-I}
\end{align}

The components of the metric can be calculated, to give
\begin{align}
g_{\sig \sig^*}
 \ &= \
 \frac{\lam}{D}
 \big( \lam e^{X} + b \big)^{\frac{1-D}{D}} e^{\wh \Psi} \; ,
 \nonumber \\
g_{\sig j^*}
 \ &= \
 \frac{\lam}{D}
 \big( \lam e^{X} + b \big)^{\frac{1-D}{D}}
 e^{\wh \Psi} \sig^* \cdot \del_{j^*} \wh \Psi \; , \\
g_{ij^*}
 \ &= \
 \frac{\lam}{D}
 \big( \lam e^{X} + b \big)^{\frac{1-D}{D}}
 e^{\wh \Psi} |\sig|^2 \cdot \del_i \wh \Psi \del_{j^*} \wh \Psi
 + \big( \lam e^{X} + b \big)^{\frac{1}{D}}
 \cdot \del_i \del_{j^*} \wh \Psi \; .
\nonumber
\end{align}
The metric of the submanifold of the surface defined by
$\sig = 0$ ($d \sig = 0$) is
\begin{align}
 g_{ij^*} \big|_{\sig = 0} (\ph, \ph^*)
 \ &= \ b^{\frac{1}{D}} \del_i \del_{j^*} \wh \Psi \; ,
\end{align}
which is the K\"{a}hler-Einstein metric
of the \kahler coset space $M$,
whose potential is given by $\wh{\Psi}$.
Therefore the total space is the canonical line bundle over $M$,
whose base manifold is endowed with
the K\"{a}hler-Einstein metric.
The total manifold can be locally written as
\begin{align}
 &{\bf R} \times G/H' \; , \ \ \
 \mbox{with } H' \ = \ H_{\rm ss.} \times U(1)^{k-1} \; .
\end{align}

Here we make a comment.
If we take the constants
in the transformation law (\ref{sigma}) as arbitrary values
instead of $h_a$,
the invariant $X'$ is different from $X$ defined in (\ref{X}).
However we obtain the same determinant with
(\ref{det}) with $X$ of (\ref{X}),
in which their difference is included in $|{\rm hol.}|^2$.
Hence (\ref{sigma}) is necessary to obtain
the ordinary differential equation (\ref{ODE}).

\section{Examples} \label{examples}

In this section, we give some examples for definiteness.
We discuss line bundles over hermitian symmetric spaces
and a non-symmetric space $SU(l+m+n)/S[U(l)\times U(m) \times U(n)]$.
Then the line bundle over ${\bf C}P^{N-1} \times {\bf C}P^{M-1}$
is considered.

\subsection{Line Bundles over Hermitian Symmetric Spaces}

The hermitian symmetric spaces (HSS) $G/H$ are
\begin{align}
& {\bf C}P^{N-1} \ = \ \frac{SU(N)}{SU(N-1) \times U(1)} \; , \ \ \
G_{N,M} \ = \ \frac{SU(N)}{SU(N-M) \times U(M)} \; , \ \ \
\frac{SO(2N)}{U(N)} \; , \ \ \
\frac{Sp(N)}{U(N)} \; , \nonumber \\
& Q^{N-2} \ = \  \frac{SO(N)}{SO(N-2) \times U(1)} \; , \ \ \
\frac{E_6}{SO(10) \times U(1)} \; , \ \ \
\frac{E_7}{E_6 \times U(1)} \; . \label{HSS}
\end{align}
The \kahler potential of HSS can be written as
\begin{align}
 \Psi \ &= \ v \log \det{}_{\eta} \xi^{\dagger} \xi \; , \ \ \
 \xi \ \in \ \GC /\wh{H} \; ,
\end{align}
with a suitable $\eta$ (see \cite{HN}).
For HSS, the constant in
the Einstein condition (\ref{Einstein}) can be calculated as
\begin{align}
 h \ &= \ \frac{1}{2 v} C_2 (G) \; ,
\end{align}
where $C_2(G)$ is the eigenvalue of the quadratic Casimir operator in
the adjoint representation of $G$
[see (\ref{h-HSS}) in Appendix \ref{ap-HSS}].
Its value is
$C_2(G)=N, N, N-1, N+1, N-2, 12$ or $18$ for
each HSS in (\ref{HSS}), respectively.
We obtain the solutions (\ref{kahler-sol}) with $X$ being
\begin{align}
 X \ &= \ \log |\sig|^2
+ \half C_2(G) \log \det{}_{\eta} \xi^{\dagger} \xi \; ,
\end{align}
which coincide with
our previous results \cite{HKN2,HKN3,HKN4},
in which we used the gauge theoretical construction.
This clarifies the origin of the coefficients in
$X$ as the coefficients in
the Einstein condition for HSS.
The total spaces can be locally regarded as
${\bf R} \times G/H'$ with $H' = H/U(1)$ in (\ref{HSS}).

In \cite{HKN2,HKN3,HKN4}, there is
the coordinate singularity in the original coordinates,
and a coordinate transformation is needed to
obtain a regular metric.
Different from these papers,
the coordinate singularity
has been avoided from beginning by the definition of
the transformation law (\ref{sigma}) of $\sigma$.

\subsection{Line Bundle over $SU(l+m+n)/S[U(l) \times U(m) \times
U(n)]$}

As an example of a line bundle over \kahler $G/H$,
which is not a HSS,
we consider $SU(l+m+n)/S[U(l) \times U(m) \times U(n)]$.
There exists two kinds of complex structures
on this coset space~\cite{BNBE}
(see Appendix \ref{G-lmn}).

I)
In one of the two inequivalent complex structures,
the complex broken generators becomes like
[see (\ref{omega-I})]
\begin{align}
 i \ph^i X_{I^*} \delta^I_i
 \ &= \ \left(
  \begin{array}{ccc}
   \mbf{0}_l & A         & B \\
           0 & \mbf{0}_m & C \\
           0 & 0         & \mbf{0}_n
  \end{array} \right) \; , \label{complex-I}
\end{align}
where $A$, $B$ and $C$ are
$l \times m$, $l \times n$ and $m
\times n$ matrices,
belonging to the
$(\mbf{l}, \ol{\mbf{m}},\mbf{1})$, $(\mbf{l}, \mbf{1},
\ol{\mbf{n}})$ and $(\mbf{1},\mbf{m},\ol{\mbf{n}})$
representations of $SU(l) \times SU(m) \times SU(n)$,
respectively.
The coset representative can be calculated as
\begin{align}
\xi \ &= \ \exp ( i \ph^i X_{I^*} \delta^I_i)
\ = \ \left(
\begin{array}{ccc}
\mbf{1}_{l} & A         & B + \half A C \\
          0 & \mbf{1}_m & C \\
          0 & 0         & \mbf{1}_n
\end{array} \right) \; . \label{xi-I}
\end{align}
Using this representative,
the \kahler potential is obtained as
\begin{align}
\Psi_{\rm I} \ &= \ \sum_{\alpha=1}^2 v_{\alpha} \log
\det{}_{\eta_\alpha} \xi^{\dagger} \xi \nonumber \\
\ &= \
v_1 \log \det_{n  \times n}
\left[ \mbf{1}_n + C\dagg C +
  \Big( B\dagg + \half C\dagg A\dagg \Big)
  \Big(B + \half A C \Big)  \right] \nonumber \\
\ & \ \ \ \
+ v_2 \log \det_{(m + n) \times (m + n)} \left(
\begin{array}{cc}
\mbf{1}_m + A\dagg A & C + A^{\dagger} ( B + \half A C ) \\
C^{\dagger} + (B^{\dagger} + \half C^{\dagger} A^{\dagger} ) A &
\mbf{1}_n + C\dagg C
 + (B\dagg + \half C\dagg A\dagg)(B + \half A C)
\end{array} \right) \; ,  \label{kahler-I}
\end{align}
where the projection operators $\eta_{\alpha}$ are given by
\begin{align}
\eta_1 \ &= \ \left(
\begin{array}{ccc}
  \mbf{0}_{l} &           &   \\
              & \mbf{0}_m &   \\
              &           & \mbf{1}_n
\end{array} \right) \; , \ \ \
\eta_2 \ = \ \left(
\begin{array}{ccc}
  \mbf{0}_{l} &           & \\
              & \mbf{1}_m & \\
              &           & \mbf{1}_n
\end{array} \right) \; .
\end{align}
The metric and the Ricci tensor at the origin
$A = B = C = 0$ can be calculated, to give\footnote{
Since the manifold is homogeneous,
it is sufficient to calculate quantities at the origin.
These can be calculated easily
using so-called \kahler normal coordinates~\cite{KNC}.
}
\bsubeq \label{ricci-metric-I}
\begin{align}
g_{ij^*}|_0 \ &= \ \left(
 \begin{array}{ccc}
 v_2 \delta_{\il \jl^*} & 0                        & 0 \\
                  0 & (v_1+v_2) \delta_{p q^*} & 0 \\
                  0 & 0                        &v_1 \delta_{u v^*}
\end{array} \right) \; , \\
\R_{ij^*}|_0 \ &= \ \left(
\begin{array}{ccc}
(l + m) \delta_{\il \jl^*} &                0 & 0 \\
         0 & (l + 2 m + n) \delta_{p q^*} & 0 \\
         0 & 0                            & (m+n) \delta_{u v^*}
\end{array} \right) \; , \label{R-I}
\end{align}
\esubeq
where we have used the indices summarized in
Table \ref{table-ABC} in Appendix \ref{ex-app}.
We thus obtain $h = (l+m)/v_2 = (m+n)/v_1 = (l+2m+n)/(v_1 + v_2)$
from the Einstein condition (\ref{Einstein}).
The line bundle over $SU(l+m+n)/S[U(l) \times U(m) \times U(n)]$
endowed with the complex structure
(\ref{complex-I}) can be obtained by the solution
(\ref{kahler-sol}) with $X$ being
\begin{align}
X \ &= \ \log |\sigma|^2 + \Psi_{\rm I} \big|_{v_1 = m+n, v_2 = l+m} \;
. \label{sol-I}
\end{align}
The manifold can be locally written as
${\bf R} \times \frac{SU(l+m+n)}{SU(l) \times SU(m) \times SU(n) \times
U(1)}$.

II)
The other complex structure of
$SU(l+m+n)/S[U(l) \times U(m) \times U(n)]$ is given by
\begin{align}
i \ph^i X_{I^*} \delta^I_i
\ &= \ \left(
 \begin{array}{ccc}
  \mbf{0}_l & 0         & B \\
          D & \mbf{0}_m & C \\
          0 & 0         & \mbf{0}_n
\end{array} \right) \; , \ \ \
\xi \ = \ \left(
\begin{array}{ccc}
\mbf{1}_{l} & 0         & B \\
          D & \mbf{1}_m & C + \half DB \\
          0 & 0         & \mbf{1}_n
\end{array} \right) \; , \label{xi-II}
\end{align}
where $B$ and $C$ are the same matrices in (\ref{xi-I}),
but $D$ is an $m \times l$ matrix,
belonging to the $(\ol{\mbf{l}},\mbf{m},\mbf{1})$
representation of
$SU(l) \times SU(m) \times SU(n)$.
The \kahler potential can be calculated as
\begin{align}
\Psi_{\rm II} \ &= \ \sum_{\alpha =1}^2 v_{\alpha} \log
\det{}_{\eta_{\alpha}} \xi^{\dagger} \xi \nonumber \\
\ &= \
v_1 \log \det_{n  \times n}
 \left[ \mbf{1}_n + B\dagg B +
  \Big( C\dagg + \half B\dagg D\dagg \Big)
  \Big( C + \half D B \Big) \right] \nonumber \\
\ & \ \ \ \
+ v_2 \log \det_{(l + n) \times (l + n)} \left(
\begin{array}{cc}
\mbf{1}_{l} + D\dagg D & B + D^{\dagger} ( C + \half D B ) \\
B^{\dagger} + (C^{\dagger} + \half B^{\dagger} D^{\dagger} ) D &
\mbf{1}_n + B\dagg B
 + (C\dagg + \half B\dagg D\dagg) (C + \half D B)
\end{array} \right) \; , \label{kahler-II}
\end{align}
where $\eta_{\alpha}$ are given by
\begin{align}
\eta_1 \ &= \ \left(
\begin{array}{ccc}
  \mbf{0}_{l} &           & \\
              & \mbf{0}_m & \\
              &           & \mbf{1}_n
\end{array} \right) \; , \ \ \
\eta_2 \ = \ \left(
\begin{array}{ccc}
  \mbf{1}_{l} &           & \\
              & \mbf{0}_m & \\
              &           & \mbf{1}_n
\end{array} \right) \; .
\end{align}
The metric and the Ricci tensor at the origin
$D = B = C = 0$ are obtained, to yield
\bsubeq \label{ricci-metric-II}
\begin{align}
 g_{ij^*}|_0 \ &= \ \left(
\begin{array}{ccc}
 v_2 \delta_{\il\jl^*} & 0                  & 0 \\
                   0 & v_1 \delta_{p q^*} & 0 \\
                   0 & 0                  &(v_1 + v_2)\delta_{u v^*}
\end{array} \right) \; , \\
\R_{ij^*}|_0 \ &= \ \left(
\begin{array}{ccc}
(l + m) \delta_{\il\jl^*} & 0             & 0 \\
             0 & (l + n) \delta_{p q^*} & 0 \\
             0 & 0                      & (2l+ m+n) \delta_{u v^*}
\end{array} \right) \; . \label{R-II}
\end{align}
\esubeq
We thus obtain $h = (l+m)/v_2 = (l+n)/v_1 = (2l+m+n)/(v_1 + v_2)$
from the Einstein condition (\ref{Einstein}).
The line bundle over $SU(l+m+n)/S[U(l) \times U(m) \times U(n)]$
endowed with the complex structure
(\ref{xi-II}) can be obtained by the solution
(\ref{kahler-sol}) with $X$ being
\begin{align}
X \ &= \ \log |\sigma|^2 + \Psi_{\rm II} \big|_{v_1 = l+n, v_2 = l+m} \;
. \label{sol-II}
\end{align}
The manifold can be also locally written as
${\bf R} \times \frac{SU(l+m+n)}{SU(l) \times SU(m) \times SU(n) \times
U(1)}$.

These two complex structures coincide in the simplest case
of $l = m = n = 1$: $G/H = SU(3)/U(1)^2$.
In this case, the manifold is locally ${\bf R} \times SU(3)/U(1)$.

\subsection{Line Bundle over ${\bf C}P^{N-1} \times {\bf C}P^{M-1}$}

As an example of the base manifold of the product of
two \kahler coset spaces, we consider
${\bf C}P^{N-1} \times {\bf C}P^{M-1}$.
The \kahler potential for the Fubini-Study metric
on this manifold is
\begin{align}
 \Psi \ &= \ v_1 \log (1 + |\phi|^2)
      + v_2 \log (1 + |\chi|^2) \; ,
\end{align}
where $\phi= \{\phi^a\}$ ($a=1,\cdots,N-1$) and
$\chi =\{\chi^{\alpha}\}$ (${\alpha}=1,\cdots,M-1$)
are coordinates of ${\bf C}P^{N-1} \times {\bf C}P^{M-1}$,
respectively.
The metric and the Ricci tensor can be calculated, to yield
\bsubeq
\begin{align}
g_{ij^*} \ &= \ \left(
 \begin{array}{cc}
 g^1_{a b^*} & 0 \\
           0 & g^2_{\alpha \beta^*}
 \end{array} \right)
\ = \ \left(
\begin{array}{cc}
 v_1 \frac{\delta_{ab} (1 + |\phi|^2) - \phi^{* a} \phi^b}{(1 +
 |\phi|^2)^2} & 0 \\
 0 & v_2 \frac{\delta_{\alpha \beta} (1 + |\chi|^2) - \chi^{* \alpha}
 \chi^{\beta}}{(1 + |\chi|^2)^2}
\end{array} \right) \; , \\
\R_{ij^*} \ &= \ \left(
\begin{array}{cc}
\frac{N}{v_1} g^1_{a b^*} & 0 \\
0 & \frac{M}{v_2} g^2_{\alpha \beta^*}
\end{array} \right) \; .
\end{align}
\esubeq
which satisfy the Einstein condition (\ref{Einstein}) with
$h_1 = N /v_1$ and $h_2 = M/ v_2$.
The quantity
\begin{align}
 X \ &= \ \log|\sig|^2 + \wh \Psi \; , \ \ \
 \wh{\Psi} \ = \ N \log (1 + |\phi|^2)
           + M \log (1 + |\chi|^2) \; , \label{CP-CPX}
\end{align}
is invariant under the transformation of
$G=U(N) \times U(M)$.
We thus have obtained the \kahler potential of
the line bundle over ${\bf C}P^{N-1} \times {\bf C}P^{M-1}$,
substituting (\ref{CP-CPX}) into
(\ref{kahler-sol}) with $D=N+M-1$.
For definiteness, the metric can be written as
\begin{align}
g_{\sig \sig^*}
 \ &= \
 \frac{\lam}{N+M-1}
 \big( \lam e^{X} + b \big)^{\frac{2-N-M}{N+M-1}} e^{\wh{\Psi}} \; ,
 \nonumber \\
g_{\sig j^*}
 \ &= \
 \frac{\lam}{N+M-1}
 \big( \lam e^{X} + b \big)^{\frac{2-N-M}{N+M-1}} e^{\wh{\Psi}} \sig^*
 \frac{V_j \ph^j}{1 + |\ph^j|^2} \; , \nonumber \\
g_{ij^*}
 \ &= \
 \frac{\lam}{N+M-1}
 \big( \lam e^{X} + b \big)^{\frac{2-N-M}{N+M-1}}
 e^{\wh{\Psi}} |\sig|^2
  \frac{V_i \ph^{* i}}{1 + |\ph^i|^2}
  \frac{V_j \ph^j}{ 1 + |\ph^j|^2 } \nonumber \\
\ & \ \ \ \
+ \big( \lam e^{X} + b \big)^{\frac{1}{N+M-1}}
   \frac{V_i \delta_{ij}}{ (1 + |\ph^i|^2)^2 } \; ,
\end{align}
where $\ph^i = \{\phi^a,\chi^{\alpha}\}$
and $V_i = N$ (or $M$) when $i$ runs over $a$ (or $\alpha$)
and we take no sum over the index $i$ or $j$.
The metric of the submanifold of $\sig=0$ is
\begin{align}
 g_{ij^*} \big|_{\sig=0} (\ph,\ph^*) \ &= \ \left(
\begin{array}{cc}
 b^{\frac{1}{N+M-1}} N (1+|\phi|^2)^{-2} & 0 \\
 0 & b^{\frac{1}{N+M-1}} M (1+|\chi|^2)^{-2}
 \end{array} \right) \; ,
\end{align}
which is the metric of ${\bf C}P^{N-1} \times {\bf C}P^{M-1}$
of definite ratio of radii.

The total space can be locally written as
${\bf R}\times \frac{SU(N) \times SU(M)}{SU(N-1)\times
SU(M-1) \times U(1)}$.
In the case of $N=M=2$, the isomorphism
$G/H = {\bf C}P^1 \times {\bf C}P^1
\simeq Q^2 = SO(4)/U(1)^2$ holds,
and the total space coincides with the conifold whose
conical singularity is resolved by $Q^2$~\cite{HKN2},
which is a special case of the solution
obtained by Pando Zayas and Tseytlin~\cite{PT}.

\section{Discussion} \label{discussion}

We have used only the Einstein condition on a base manifold,
but not the symmetry as in \cite{HKN1,HKN2,HKN3,HKN4}.
This implies the existence of
a Calabi-Yau metric on a canonical line bundle
on a base manifold without any isometry,
but the only known K\"{a}hler-Einstein spaces
with positive scalar curvature are
\kahler coset spaces \cite{Be} discussed in this paper.
On the other hand, a Calabi-Yau metric of
cohomogeneity one which is not in a form of a bundle
can be constructed~\cite{HKN1}
as the Stenzel metric~\cite{St}.
In terms of the nonlinear realization,
the total space in this paper can be understood
by the existence of the so-called quasi-Nambu-Goldstone
(QNG) boson~\cite{Ni1}.
If there is only one QNG boson in generic points
of the manifold, it is cohomogeneity one.
The generalization to a vector bundle can be also considered,
using the matter coupling method
in the nonlinear realization~\cite{BKMU,matter}.
We consider that almost all Calabi-Yau manifolds
of cohomogeneity one
can be constructed using these methods.
Explicit metrics of Calabi-Yau manifolds
would provide more precise information
about superstring propagating on these manifolds
than topological structures.

\section*{Acknowledgements}

We would like to thank Y.~Sakane for useful comments.
This work was supported in part by
the Grant-in-Aid for Scientific Research.
The work of M.N. was supported by the U.S.
Department of Energy under grant DE-FG02-91ER40681 (Task B).


\begin{appendix}

\section*{Appendix}

\section{Ricci Tensor of Riemann and \kahler Coset Spaces}
\label{tensors}
In this appendix we calculate Ricci tensors
of Riemann and \kahler coset spaces.
Let us first consider Riemann coset spaces $G/H$~\cite{CRW,Sh}.
After we construct the Ricci tensor by using the structure constants of
$G$,
we require the coset to be \kahler.

\subsection{Riemann Coset Spaces}
Here we assume the group $G$ is semi-simple.
We prepare a set of generators $S_{\wh{a}}$
spanning the algebra ${\cal H}$
and a set of generators $X_{\wh{I}}$
generating the coset $G/H$:
\bsubeq
\begin{align}
{\cal H} \ &= \ \{ S_{\wh{a}} \} \ \ \
  \mbox{: unbroken generators} \; , \label{gen-H} \\
{\cal G} - {\cal H} \ &= \ \{ X_{\wh{I}} \} \ \ \
  \mbox{: broken generators} \; . \label{gen-G/H}
\end{align}
\esubeq
We consider reductive coset spaces $G/H$:
$[X_{\wh{I}}, S_{\wh{a}}] \propto X_{\wh{J}}$,
in which case $X_{\wh{I}}$ belong to, in general, reducible
representations of $H$.
The algebra can be written as
\bsubeq
\begin{align}
[ X_{\wh{I}} , X_{\wh{J}} ] \ &= \ i f_{\wh{I} \wh{J}}{}^{\wh{K}}
X_{\wh{K}} + i f_{\wh{I} \wh{J}}{}^{\wh{a}} S_{\wh{a}} \; ,
\label{alg-1} \\
[ X_{\wh{I}} , S_{\wh{a}} ] \ &= \ i f_{\wh{I} \wh{a}}{}^{\wh{J}}
X_{\wh{J}} \; , \label{alg-2} \\
[ S_{\wh{a}} , S_{\wh{b}} ] \ &= \ i f_{\wh{a} \wh{b}}{}^{\wh{c}}
S_{\wh{c}} \; . \label{alg-3}
\end{align}
\esubeq
Here we have taken generators as hermitian.
When there exists an automorphism, defined by
$S \to S$ and $X \to - X$,
a coset space is called a symmetric space
and $f_{\wh{I} \wh{a}}{}^{\wh{J}}$ vanishes.

Using real coordinates $\ph^i$,
the coset representative of $G/H$ can be written as
$U(\ph) = \exp (i \ph^i X_{\wh{I}} \delta^{\wh{I}}_i)$.
The ${\cal G}$-valued 1-form,
\beq
 \alpha(\ph) \ \equiv \ \1{i} U^{-1} d U \,,
\eeq
called the Maurer-Cartan 1-form, can be decomposed like
\begin{align}
\alpha \ &= \ \wt{e}^{\wh{I}}(\ph) X_{\wh{I}}
 + \om^{\wh{a}} (\ph) S_{\wh{a}}
\ = \
 \big[ \wt{e}^{\wh{I}}_i (\ph) X_{\wh{I}}
  + \om^{\wh{a}}_i (\ph) S_{\wh{a}} \big] d\ph^i
\; ,  \label{MC-1}
\end{align}
where $\om^{\wh{a}}_i$ is called a $H$-connection
and $\wt{e}^{\wh{I}}_i$ is a (rescaled) vielbein.
The most general $G$-invariant metric
$g_{ij}(\ph)$ on $G/H$
contains real positive scale constants ${r_I}$
associated with
$H$-irreducible representations of broken generators
$X_{\wh{I}}$
($r_I = r_J$ if $X_{\wh{I}}$ and $X_{\wh{J}}$ belong to
the same representation of $H$):
\begin{align}
 g_{ij}(\ph)
 \ = \ ({r_I})^{-2}
    \delta_{\wh{I} \wh{J}} \,
     \wt{e}^{\wh{I}}_i(\ph) \wt{e}^{\wh{J}}_j(\ph)
 \ = \ \delta_{\wh{I} \wh{J}}\,
       e^{\wh{I}}_i(\ph) e^{\wh{J}}_j(\ph) \; .
 \label{G-metric}
\end{align}
Here the metric of the tangent space is normalized
in the ordinary basis of $e^{\wh{I}}$,
defined
by rescaling $e^{\wh{I}} = ({r_I})^{-1} \wt{e}^{\wh{I}}$.
They can be expanded in the coset coordinates, like
\beq
 \wt{e}^{\wh{I}}_i (\ph)
   \ = \ \delta^{\wh{I}}_i + O(\ph) \;, \ \ \
 e^{\wh{I}}_i (\ph)
   \ = \ (r_I)^{-1} \delta^{\wh{I}}_i + O(\ph) \, .
 \label{viel-exp}
\eeq

The Maurer-Cartan 1-form satisfies an equation,
$d \alpha = - i \alpha \w \alpha$,
called the Maurer-Cartan equation.
Under the decomposition (\ref{MC-1}), this equation becomes
\begin{align}
d \wt{e}^{\wh{I}} \ &= \
\half f_{\wh{J} \wh{K}}{}^{\wh{I}} \wt{e}^{\wh{J}} \w \wt{e}^{\wh{K}}
+ f_{\wh{K} \wh{b}}{}^{\wh{I}} \wt{e}^{\wh{K}} \w \om^{\wh{b}} \; , \ \ \
d \om^{\wh{a}} \ = \
\half f_{\wh{J} \wh{K}}{}^{\wh{a}} \wt{e}^{\wh{J}} \w \wt{e}^{\wh{K}}
+ \half f_{\wh{b} \wh{c}}{}^{\wh{a}} \om^{\wh{b}} \w
\om^{\wh{c}} \; . \label{vielbein-1}
\end{align}
In ordinary basis of vielbeins,
these equations can be rewritten as
\begin{align}
d e^{\wh{I}} \ &= \ \half \frac{{r_J} {r_K}}{{r_I}} f_{\wh{J}
 \wh{K}}{}^{\wh{I}} e^{\wh{J}} \w e^{\wh{K}} + \frac{{r_K}}{{r_I}}
 f_{\wh{K} \wh{b}}{}^{\wh{I}} e^{\wh{K}} \w \om^{\wh{b}} \; , \ \ \
d \om^{\wh{a}} \ =
 \ \half {{r_J} {r_K}} f_{\wh{J} \wh{K}}{}^{\wh{a}}
 e^{\wh{J}} \w e^{\wh{K}} + \half f_{\wh{b} \wh{c}}{}^{\wh{a}}
 \om^{\wh{b}} \w \om^{\wh{c}} \; . \label{vielbein-2}
\end{align}
The connection 1-form $\Om^{\wh{I}}{}_{\wh{J}}$
and the curvature 2-form $R^{\wh{I}}{}_{\wh{J}}$
can be read off from
the Cartan's structure equations:
\begin{align}
d e^{\wh{I}} \ &= \ - \Om^{\wh{I}}{}_{\wh{J}} \w e^{\wh{J}} \; , \ \ \
R^{\wh{I}}{}_{\wh{J}} \ = \
d \Om^{\wh{I}}{}_{\wh{J}}
+ \Om^{\wh{I}}{}_{\wh{K}} \w \Om^{\wh{K}}{}_{\wh{J}}
\ = \
\half R^{\wh{I}}{}_{\wh{J} \wh{L} \wh{M}} e^{\wh{L}} \w e^{\wh{M}} \;
, \label{Cartan}
\end{align}
where we have imposed
the torsion-free condition, and $R^{\wh{I}}{}_{\wh{J}
\wh{L} \wh{M}}$ is the Riemann tensor. From equations
(\ref{vielbein-2}) and (\ref{Cartan}),
we can write down the Riemann tensor of
the coset space $G/H$ endowed with the $G$-invariant metric
(\ref{G-metric}):
\bsubeq
\begin{align}
R^{\wh{I}}{}_{\wh{J} \wh{L} \wh{M}} \ &= \
{r_L} {r_M}
f_{\wh{J} \wh{b}}{}^{\wh{I}} f_{\wh{L} \wh{M}}{}^{\wh{b}}
+ \half \frac{{r_L} {r_M}}{{r_K}} \scale{I}{J}{K}
f_{\wh{J} \wh{K}}{}^{\wh{I}} f_{\wh{L} \wh{M}}{}{\wh{K}} \nonumber \\
\ & \ \ \ \
+ \frac{1}{4} \scale{I}{K}{L} \scale{K}{J}{M}
f_{\wh{J} \wh{M}}{}^{\wh{K}} f_{\wh{K} \wh{L}}{}^{\wh{I}}
- \frac{1}{4} \scale{I}{K}{M} \scale{K}{J}{L}
f_{\wh{J} \wh{L}}{}^{\wh{K}} f_{\wh{K} \wh{M}}{}^{\wh{I}}
\; , \label{riemann-1} \\
\scale{I}{J}{K} \ &\equiv \
\frac{{r_J} {r_K}}{{r_I}} +  \frac{{r_I} {r_K}}{{r_J}} -
\frac{{r_I} {r_J}}{{r_K}} \; .
\end{align}
\esubeq
The Ricci tensor
$\R_{\wh{J} \wh{M}} \equiv R^{\wh{I}}{}_{\wh{J} \wh{I} \wh{M}}$
can be calculated, to yield
\begin{align}
\R_{\wh{J} \wh{M}} \ &\equiv \
{r_J} {r_M} f_{\wh{J} \wh{b}}{}^{\wh{I}} f_{\wh{I} \wh{M}}{}^{\wh{b}}
+ \frac{1}{4}
f_{\wh{J} \wh{K}}{}^{\wh{I}} f_{\wh{I} \wh{M}}{}^{\wh{K}}
\scale{M}{K}{I} \scale{J}{I}{K}
\; .
\end{align}

\subsection{\kahler Coset Spaces}

Now let us consider the \kahler coset spaces
imposing conditions on the Riemann coset spaces
following Ref.~\cite{IKK}.
We rewrite the broken and unbroken generators as
\bsubeq
\begin{align}
{\cal H} \ &= \ \{ S_{\wh{a}} \} \ = \ \{ S_a , Y_{\alpha} \} \; , &&
\mbox{: unbroken generators} \; , \label{kahler-H} \\
{\cal G} - {\cal H} \ &= \ \{ X_{\wh{I}} \} \ = \ \{ X_I ,
X_{I^*} \} \; , && \mbox{: broken generators} \;
. \label{kahler-G/H}
\end{align}
\esubeq
Here $X_I$ and $X_{I^*}$ are non-hermitian
and hermitian conjugate to each other:
$X_{I^*} = (X_I)^{\dagger}$,
$S_a$ generate the semi-simple subgroup $H_{\rm ss.}$ of $H$,
and $Y_{\alpha}$ is a generator commuting
with any generators of ${\cal H}$,
that is,
$Y_{\alpha}$ is a $U(1)$ generator of the torus in $H$.
We assume that $Y_{\alpha}$ is orthogonal to each other:
$\tr (Y_{\alpha} Y_{\beta})
= N_{\alpha} \delta_{\alpha \beta}$,
with a normalization constant $N_{\alpha}$.
In this subsection, we denote holomorphic
coordinates by $\ph^i$, and replace real coordinates
in the last subsection
by $\ph^i$ and $\ph^{*i}$. Hence, for instance,
vielbein is not holomorphic: $e^I(\ph,\ph^*)
= e^I_i(\ph,\ph^*) d\ph^i$
[$e^{I^*}(\ph,\ph^*)
= e^{I^*}_{i^*}(\ph,\ph^*) d\ph^{*i}$].

We define the $\mbf{Y}$-charge by
\begin{align}
\mbf{Y} \ &\equiv \ \sum_{\alpha=1}^k v_{\alpha} Y_{\alpha} \; ,
\label{Y-gen}
\end{align}
where $v_{\alpha}$ is a real positive constant and
$k$ is the dimension of the torus of $H$ ($1 \leq k \leq \dim G$).
The generators with the negative $\mbf{Y}$-charges are defined
as the complex broken generators,
generating the complex coset space $\GC/\wh{H}$~\cite{IKK}:
the $\mbf{Y}$-charges of the generators, $X_I$ and $X_{I^*}$,
and ${\cal H}$ are
\begin{align}
[ \mbf{Y}, X_I ] \ &= \ x_I X_I \; , \ \ \
[ \mbf{Y}, X_{I^*} ] \ = \ - x_I X_{I^*} \; , \ \ \
[ \mbf{Y}, {\cal H} ] \ = \ 0 \; , \label{Y-charge}
\end{align}
where $x_I$ is a real positive constant related to
$r_I$ by $x_I = (r_I)^{-2}$. From these commutation relations,
the structure constants are constrained by
\begin{align}
&f_{IJ}{}^{K^*} \ = \ f_{I^* J^*}{}^{K} \ = \
f_{IJ}{}^{\wh{a}} \ = \ f_{I^* J^*}{}^{\wh{a}} \ = \
f_{I \wh{a}}{}^{J^*} \ = \ f_{I^* \wh{a}}{}^{J} \ = \
f_{IK}{}^I \ = \ f_{I^* K^*}{}^{I^*} \ = \ 0 \; .
\label{str-a}
\end{align}
We would like to obtain further constraints from
the \kahler condition:
the \kahler form $\Om$ defined by
\begin{align}
 \Om \ &= \
  i g_{ij^*} d \ph^{i} \w d \ph^{*j}
  \ = \ i \delta_{I J^*} e^{I} \w e^{J^*}
  \ = \ i x_I \delta_{I J^*} \wt{e}^{I} \w \wt{e}^{J^*} \;
\label{kahler-form}
\end{align}
must be closed: $d \Om = 0$.
This condition gives
\bsubeq
\begin{align}
f_{IJ}{}^K \ &\neq \ 0 \; , \ \ \ f_{I^* J^*}{}^{K^*} \
\neq \ 0 && \mbox{if} \ \ x_I + x_J \ = \ x_K \; , \label{str-b} \\
f_{I J^*}{}^K \ &\neq \ 0 && \mbox{if} \ \ x_I - x_J \ = \
x_K \; , \label{str-c} \\
f_{I J^*}{}^{K^*} \ &\neq \ 0 && \mbox{if} \ \ x_I -
x_J \ = \ - x_K \; , \label{str-d} \\
f_{I \wh{a}}{}^J \ &\neq \ 0 \; , \ \ \ f_{I^* \wh{a}}{}^{J^*}
\ \neq 0 && \mbox{if} \ \ x_I \ = \ x_J \; . \label{str-e}
\end{align}
\esubeq
Using these constraints,
we obtain the Ricci tensor of \kahler coset spaces $G/H$ as
\begin{align}
\R_{J M^*} \ &= \
{r_J} {r_M} f_{J \wh{b}}{}^I f_{I M^*}{}^{\wh{b}}
+ \frac{1}{4} \Big( f_{J K}{}^I f_{I M^*}{}^K
+  f_{J K^*}{}^I f_{I M^*}{}^{K^*}
+ f_{J K^*}{}^{I^*} f_{I^* M^*}{}^{K^*} \Big)
\scale{M}{K}{I} \scale{J}{I}{K} \; . \label{ricci-3}
\end{align}

The Einstein condition (\ref{Einstein}) in the vielbein basis is
\beq
 \R_{IJ^*} \ = \ h \delta_{IJ^*} \, . \label{Einstein2}
\eeq
We explicitly solve this equation for some examples
in the next section.


\section{Examples} \label{ex-app}

In this appendix we consider two examples of
the \kahler $G/H$.
We discuss the hermitian symmetric spaces (HSS)~\cite{HKN4}
followed by a non-symmetric space
$SU(l+m+n)/S[U(l) \times U(m) \times U(n)]$~\cite{IKK}.
We see that the HSS are Einstein;
on the other hand, the Einstein condition
on the non-symmetric space
reduces to algebraic equations among
scale constants in the metric.

\subsection{Hermitian Symmetric Spaces} \label{ap-HSS}

In this subsection let us consider symmetric spaces.
The HSS are K\"{a}hler,
since the \kahler condition is automatically satisfied.

The HSS have
constraints on the structure constants given by
$f_{IJ}{}^K = f_{I^* J^*}{}^{K^*} = f_{I
J^*}{}^K = f_{I J^*}{}^{K^*} = 0$,
and each metric contains only one scale constant:
$r_I \equiv r$ ($x_I = r^{-2} \equiv v$) for all $I$.
Therefore the Ricci tensor (\ref{ricci-3}) is reduced to
\begin{align}
\R_{J M^*} \ &= \
 r^2 f_{J \wh{b}}{}^I f_{I M^*}{}^{\wh{b}} \ = \
\frac{r^2}{2} \big( f_{J \wh{b}}{}^I f_{I M^*}{}^{\wh{b}}
+ f_{J I^*}{}^{\wh{b}} f_{\wh{b} M^*}{}^{I^*} \big)
\ = \ - \frac{r^2}{2}
f_{\wh{A} J}{}^{\wh{B}} f_{\wh{B} M^*}{}^{\wh{A}}
\nonumber \\
\ &= \  \frac{r^2}{2}
\tr \big[ {\rm ad.} (X_J) {\rm ad.} (X_{M^*}) \big]
\ = \ \frac{1}{2 v} C_2 (G) \delta_{J M^*} \; .
 \label{HSS-Ricci}
\end{align}
where the index $\wh{A}$ represents the whole generators
of ${\cal G}$: $\wh{A} \in \{ I, I^*, \wh{a}\}$.
Here we have adopted the adjoint representation $[{\rm
ad.}(X_I)]_{\wh{A}}{}^{\wh{B}} = i f_{\wh{A} I}{}^{\wh{B}}$,
and defined $C_2 (G)$ as
the eigenvalue of the quadratic Casimir
operator in the adjoint representation of $G$.

~From (\ref{HSS-Ricci}),
the HSS are Einstein (\ref{Einstein2}) in which $h$ is given by
\begin{align}
 h \ = \frac{1}{2 v} C_2 (G) \; . \label{h-HSS}
\end{align}

\subsection{A Non-symmetric Space
$SU(l+m+n)/S[U(l) \times U(m) \times U(n)]$} \label{G-lmn}

In this subsection,
let us consider the coset space
$G/H = SU(l+m+n)/S[U(l)
\times U(m) \times U(n)]$, which is non-symmetric.

\subsubsection{Complex Structures}
First we discuss the complex structures of the coset space
$SU(l+m+n)/S[U(l) \times U(m) \times U(n)]$.
This coset admits two kinds of inequivalent complex
structures~\cite{IKK,BNBE}.
The unbroken group $H = SU(l) \times SU(m) \times SU(n) \times U(1)^2$
has two $U(1)$ generators $Y_1$ and $Y_2$,
which can be chosen as
\begin{align}
Y_1 \ &= \ \left(
\begin{array}{ccc}
 n \mbf{1}_{l} &           & \\
               & \mbf{0}_m & \\
               &           & -l \mbf{1}_n
\end{array} \right) \; , \ \ \
Y_2 \ = \ \left(
\begin{array}{ccc}
 m \mbf{1}_{l} &                  & \\
               & -(l+n) \mbf{1}_m & \\
               &                  & m \mbf{1}_n
\end{array} \right) \; , \label{Y1-Y2}
\end{align}
embedded into
the fundamental representation of $G = SU(l+m+n)$.
These generators are orthogonal to each other:
$\tr (Y_{\alpha} Y_{\beta}) = N_{\alpha} \delta_{\alpha \beta}$.
The whole generators of ${\cal SU}(l+m+n)$ can be
decomposed as
\begin{align}
\ \left(
\begin{array}{ccc}
 {\cal SU}(l) & A            & B \\
      \bar{A} & {\cal SU}(m) & C \\
      \bar{B} & \bar{C}      & {\cal SU}(n)
\end{array} \right) \; . \label{decomposition}
\end{align}
Here ${\cal SU}(l)$, ${\cal SU}(m)$ and ${\cal SU}(n)$
are the blocks of the unbroken subalgebra ${\cal H}$,
and $A$, $B$ and $C$ are blocks of the broken generators
belonging to irreducible representations of $H$
(we simply call them ``irreducible blocks'').
We explain these irreducible blocks:
the irreducible block $A$ consists of
$l \times m$ generators denoted by $X_{\il}$
($\il=1,\cdots,lm$).
These generators belong to
the $(\mbf{l}, \ol{\mbf{m}}, \mbf{1})$-representation
of $SU(l)\times SU(m)\times SU(n)$,
with their $Y_1$- and $Y_2$-charges being
$n$ and $l+m+n$, respectively.
The block $\bar{A}$ is hermitian conjugate of $A$.
Our notation and $H$-representations of
irreducible blocks $A$, $B$ and $C$ are
summarized in Table \ref{table-ABC}.

\begin{table}[h]
\begin{center}
\begin{tabular}{c||c|c|c}
irreducible block & $A$ & $B$ & $C$ \\ \hline\hline
index & $\il,\jl,\cdots$ & $p,q,\cdots$ & $u, v, \cdots$ \\
matrix size & $l \times m$ & $l \times n$ & $m \times n$ \\
$H$-representations
&
$(\mbf{l}, \ol{\mbf{m}}, \mbf{1})^{n,l+m+n}$ & $(\mbf{l}, \mbf{1},
\ol{\mbf{n}})^{l+n,0}$ & $(\mbf{1}, \mbf{m},
\ol{\mbf{n}})^{l,-(l+m+n)}$
\end{tabular}

\caption{The broken generators of $SU(l+m+n)/S[U(l) \times U(m)
\times U(n)]$.
We denote complex broken generators
belonging to $A$, $B$ or $C$
by $X_{\il}$, $X_p$ or $X_u$.
The $H$-representations are denoted by
the representations of $SU(l) \times SU(m) \times SU(n)$
in the braces
and the $Y_1$- and $Y_2$-charges as the indices.
}
\label{table-ABC}
\end{center}
\end{table}

Depending on real constants $v_1$ and $v_2$ in
the definition (\ref{Y-gen}) of $\mbf{Y}$-charge,
the complex structure of $G/H$ continuously varies.
There exist, however, two classes of inequivalent
complex structures, which are not continuously
deformed to each other, represented, for instance,
by~\cite{BNBE}
\begin{align}
{\rm I)}\;\;\; \mbf{Y} \ &= \ - Y_1  \; , \ \ \ \non
{\rm II)}\;\;\; \mbf{Y} \ &= \ -m Y_1 + n Y_2  \; .
\end{align}
I) In the case of $\mbf{Y} = - Y_1$,
the broken generators $X_{I^*}$ of ${\cal G}^{\bf C} -
\wh{{\cal H}}$ with negative $\mbf{Y}$-charges are given by
\begin{align}
{\cal G}^{\bf C} - \wh{{\cal H}} \ &= \  \{ A, B, C \} \; ,
\label{omega-I}
\end{align}
and the scale parameters $x_I$ are related as
\begin{align}
x_A + x_C \ &= \ x_B \; . \label{constraint-OmI}
\end{align}
II) In the case of $\mbf{Y} =  -m Y_1 + n Y_2$,
the broken generators and the relation among
scale parameters are given by
\begin{align}
{\cal G}^{\bf C} - \wh{{\cal H}} \ &= \  \{ \bar{A}, B, C \} \; ,
\ \ \
x_A + x_B \ = \ x_C \; . \label{constraint-OmII}
\end{align}

\subsubsection{Ricci Tensor and the Einstein Condition}
We show that the Einstein condition reduces to
a set of algebraic equations
in cases of \kahler coset spaces.
To this end,
let us calculate the Ricci tensor in the both cases.\\
I) In the case of $\mbf{Y} = - Y_1$,
the non-zero components of the Ricci tensor are given by
\bsubeq \label{ricci-I}
\begin{align}
\R_{\il \jl^*} \ &= \
 r_{\il} r_{\jl} f_{\il \wh{a}}{}^{\kl} f_{\kl \jl^*}{}^{\wh{a}}
 + \frac{r_{\il} r_{\jl}}{4}
 \big(
  f_{\il u}{}^p f_{p \jl^*}{}^u
  + f_{\il p^*}{}^{u^*} f_{u^* \jl^*}{}^{p^*}
 \big) \scale{A}{B}{C} \scale{A}{C}{B} \; , \\
\R_{p q^*} \ &= \
 r_p r_q f_{p \wh{a}}{}^{r} f_{r q^*}{}^{\wh{a}}
 + \frac{r_p r_q}{4} \big(
   f_{p \il^*}{}^u f_{u q^*}{}^{\il^*}
 + f_{p u^*}{}^{\il} f_{\il q^*}{}^{u^*}
 \big) \scale{B}{A}{C} \scale{B}{C}{A} \; , \\
\R_{u v^*} \ &= \
r_u r_v f_{u \wh{a}}{}^{w} f_{w v^*}{}^{\wh{a}}
+ \frac{r_u r_v}{4} \big(
 f_{u \il}{}^p f_{p v^*}{}^{\il}
 + f_{u p^*}{}^{\il^*} f_{\il^* v^*}{}^{p^*}
\big) \scale{C}{A}{B} \scale{C}{B}{A} \; .
\end{align}
\esubeq
Due to the constraint (\ref{constraint-OmI}),
we have
\begin{align}
\scale{A}{B}{C} \scale{A}{C}{B} \ &= \ \scale{C}{A}{B} \scale{C}{B}{A}
\ = \ 0 \; , \ \ \
\scale{B}{A}{C} \scale{B}{C}{A} \ = \ \frac{4}{x_B} \; . \label{scale-I}
\end{align}

In order to evaluate the various terms in (\ref{ricci-I}),
we use the normalization of generators
$T_{\wh{A}}$ of $SU(N)$ in the
fundamental representation, given by
\begin{align}
\tr (T_{\wh{A}} T_{\wh{B}}) \ &\equiv \ \delta_{\wh{A} \wh{B}} \;
. \label{normalization}
\end{align}
In this normalization,
the eigenvalues of the quadratic Casimir operators
$C_2 (N)$ and $C_2 ({\rm ad.})$
in the fundamental and the adjoint representations,
respectively, are given by
\begin{align}
C_2 (N) \ &= \ \frac{N^2 - 1}{N} \; , \ \ \
C_2 ({\rm ad.}) \ = \ 2 N \; . \label{casimir-value}
\end{align}
We also fix the normalization
the $U(1)$ generators $Y_{\alpha}$ in the same way.

Using these equations,
we can calculate the structure constants in
(\ref{ricci-I}) explicitly.
From the algebra (\ref{alg-2}), for instance,
we can obtain the following relation:
\begin{align}
 f_{\il \wh{a}}{}^{\kl} f_{\kl \jl^*}{}^{\wh{a}}
 \ &= \ C_2 (A) \delta_{\il \jl^*}
\; , \ \ \ \mbox{etc.}
\end{align}
Here $C_2 (A)$ is the eigenvalue of the quadratic Casimir operator
of the irreducible block $A$ in (\ref{decomposition}),
which can be calculated,
using the relations (\ref{casimir-value}), to give
\begin{align}
C_2 (A) \ &= \ \frac{l^2-1}{l} + \frac{m^2-1}{m} +
\frac{n}{l(l+n)} + \frac{l+m+n}{m(l+n)} \ = \ l + m \;
. \label{C2A}
\end{align}
In the same way, we obtain
\begin{align}
f_{p \wh{a}}{}^{r} f_{r q^*}{}^{\wh{a}} \ &= \ C_2 (B) \delta_{p q^*}
\ = \ (l + n) \delta_{p q^*} \; , \ \ \
f_{u \wh{a}}{}^{w} f_{w v^*}{}^{\wh{a}} \ = \ C_2 (C) \delta_{u v^*}
\ = \ (m + n) \delta_{u v^*} \; . \label{C2-BC}
\end{align}
Moreover, the fact of $G=SU(N)$ provides us the relations,
given by
\begin{align}
 f_{\il u}{}^p f_{p \jl^*}{}^u \ &= \ n \delta_{\il \jl^*} \; , \ \ \
 f_{p \il^*}{}^u f_{u q^*}{}^{\il^*}
   \ = \ m \delta_{p q^*} \; , \ \ \
 f_{u \il}{}^p f_{p v^*}{}^{\il}
   \ = \ l \delta_{u v^*} \;.
 \label{second-terms}
\end{align}
Therefore the Ricci tensor (\ref{ricci-I}) can be
explicitly calculated, to yield
\begin{align}
\R_{I J^*} \ &= \ \left(
\begin{array}{ccc}
\frac{1}{x_A} (l+m) \delta_{\il \jl^*} & 0 & 0 \\
0 & \frac{1}{x_B} (l + 2 m + n) \delta_{p q^*} & 0 \\
0 & 0 & \frac{1}{x_C} (m+n) \delta_{u v^*}
\end{array} \right) \; . \label{ricci-I-2}
\end{align}
This concides with the direct calculation
(\ref{R-I}) at the origin due to (\ref{viel-exp}).

The Einstein condition (\ref{Einstein2})
reduces to a set of algebraic equations given by
\begin{align}
h \ &= \ \frac{l + m}{x_A} \ = \ \frac{l + 2m + n}{x_B} \ = \
\frac{m+n}{x_C} \; . \label{app-I-sol}
\end{align}

~\\
II)
In the case of $\mbf{Y} = - m Y_1 + n Y_2$,
the components of the Ricci tensor are given by
\bsubeq \label{ricci-II}
\begin{align}
\R_{\il \jl^*} \ &= \
 r_{\il} r_{\jl} f_{\il \wh{a}}{}^{\kl} f_{\kl \jl^*}{}^{\wh{a}}
 + \frac{r_{\il} r_{\jl}}{4} \big(
     f_{\il p}{}^u f_{u \jl^*}{}^p
   + f_{\il u^*}{}^{p^*} f_{p^* \jl^*}{}^{u^*}
  \big) \scale{A}{B}{C} \scale{A}{C}{B} \; , \\
\R_{p q^*} \ &= \
 r_p r_q f_{p \wh{a}}{}^{r} f_{r q^*}{}^{\wh{a}}
+ \frac{r_p r_q}{4} \big(
    f_{p \il}{}^u f_{u q^*}{}^{\il}
  + f_{p u^*}{}^{\il^*} f_{\il^* q^*}{}^{u^*}
\big) \scale{B}{A}{C} \scale{B}{C}{A} \; , \\
\R_{u v^*} \ &= \
r_u r_v f_{u \wh{a}}{}^{w} f_{w v^*}{}^{\wh{a}}
+ \frac{r_u r_v}{4} \big(
   f_{u \il^*}{}^p f_{p v^*}{}^{\il^*}
 + f_{u p^*}{}^{\il} f_{\il v^*}{}^{p^*}
\big) \scale{C}{A}{B} \scale{C}{B}{A} \; .
\end{align}
\esubeq
By the same discussion from (\ref{scale-I}) to (\ref{C2-BC}),
we obtain the Ricci tensor, given by
\begin{align}
\R_{I J^*} \ &= \ \left(
\begin{array}{ccc}
\frac{1}{x_A} (l+m) \delta_{\il \jl^*} & 0 & 0 \\
  0 & \frac{1}{x_B} (l +n) \delta_{p q^*} & 0 \\
  0 & 0 & \frac{1}{x_C} (2l + m+n) \delta_{u v^*}
\end{array} \right) \; . \label{ricci-II-2}
\end{align}
This also concides with the direct calculation
(\ref{R-II}) at the origin due to (\ref{viel-exp}).

The Einstein condition (\ref{Einstein2}) reduces to
a set of algebraic equations:
\begin{align}
h \ &= \ \frac{l + m}{x_A} \ = \ \frac{l + n}{x_B} \ = \
\frac{2l+m+n}{x_C} \; . \label{app-II-sol}
\end{align}

\end{appendix}


\end{document}